\documentclass[11pt]{article}

\usepackage[dvips]{epsfig}
\usepackage{epsfig}
\usepackage{graphicx}
\usepackage{color}
\usepackage{fullpage}

\usepackage{latexsym}
\usepackage{amsmath}
\usepackage{amssymb}

\long\def\commabs #1\commabsend{}

\newcommand{\reals}{\hbox{$\rlap{\rm I} \> \kern-.2mm{\rm R}$}}

\newcommand{\etal}{{\em et al.\ }}

\commabs

\commabsend

\newcommand{\dist}{\mbox{\bf dist}}

\newcommand{\eps}{\epsilon}

\def\inline#1:{\par\vskip 7pt\noindent{\bf #1:}\hskip 10pt}
\def\Proof{\par\noindent{\bf Proof:~}}

\newtheorem{theorem}{Theorem}[section]
\newtheorem{lemma}[theorem]{Lemma}

\newtheorem{claim}{Claim}[section]

\newtheorem{definition}{Definition}

\newcommand{\estimate}{\mbox{\bf Estimate}}
\newcommand{\checkind}{\mbox{\bf Check-Ind}}
\newcommand{\findleg}{\mbox{\bf Find-Legitimate}}

\def\inline#1:{\par\vskip 7pt\noindent{\bf #1:}\hskip 10pt}
\def\Proof{\par\noindent{\bf Proof:~}}
\def\blackslug{\hbox{\hskip 1pt \vrule width 4pt height 8pt
    depth 1.5pt \hskip 1pt}}

\def\QED{\quad\blackslug\lower 8.5pt\null\par}
\renewcommand{\paragraph}[1]{\par\noindent\textbf{#1}}

\def\beginsmall#1{\vspace{-\parskip}\begin{#1}\itemsep-\parskip}
\def\endsmall#1{\end{#1}\vspace{-\parskip}}

\begin{document}
%\title{Approximate Distance Oracle with Even Faster Query Time: $2k-1$ stretch with constant query time}
\title{Approximate Distance Oracle with Constant Query Time}

%\author{
%Shiri Chechik\inst{1}}
%
%\institute
%Department of Computer Science, The Weizmann Institute, Rehovot, Israel\\
%\email{\{shiri.chechik,david.peleg\}@weizmann.ac.il}
%}

\author{Shiri Chechik
\thanks{Microsoft Research Silicon Valley, Mountain View CA, USA.
Email: {\tt schechik@microsoft.com}.}}

\date{\today}
\maketitle

\thispagestyle{empty}

\begin{abstract}
%Given a graph $G =(V,E)$,
An approximate distance oracle is a succinct data structure that provides fast answers to distance queries between any two nodes.

In this paper we consider approximate distance oracles for general undirected graphs with non-negative edge weights with constant query time.
We present a distance oracle of size $O(k n^{1+1/k})$, with $2k-1$ stretch and $O(1)$ query time.
This improves the $O(\log{k})$ query time of Wulff-Nilsen's distance oracle [SODA '13], which in turn improved the $O(k)$ query time of Thorup and Zwick's distance oracle [J. ACM '05].
 \end{abstract}
%
%\begin{keyword}
%forbidden sets \sep fault-tolerance \sep compact routing
%\end{keyword}

\maketitle

%==========================================================================
\section{Introduction}

Finding shortest paths is perhaps one of the most fundamental and studied computational problems.
A great number of papers deal with different variants of this problem.
For example, the well known Dijkstra and Bellman-Ford algorithms allow computing the shortest path distance between any pair of nodes.
In many applications it is desirable to retrieve shortest path distances extremely fast, ideally in time that is independent of the network size.

A distance oracle is a data structure that allows fast retrieval of a distance estimate for any pair of nodes.
A naive solution to accomplish this is to invoke an all pairs shortest paths algorithm and store the distance matrix.
Using the pre-computed distance matrix, distance queries can be answered in constant time.
The main disadvantages of this solution is that the space may be too large (quadratic in the number of nodes) and that computing all pairs shortest paths may take too long.
To overcome these drawbacks, much of the work on distance oracles considers approximated distances.

The distance oracle is said to be of stretch $k$ (or $k$-approximate distance oracle), if for every two vertices $s$ and $t$, the distance $\widehat{\dist}(s,t)$ returned by the distance oracle satisfies $\dist(s,t) \leq  \widehat{\dist}(s,t) \leq k \cdot \dist(s,t)$.

The focus on designing distance oracles is often on the tradeoff between several parameters:
the construction time (the time it takes to construct the distance oracle), the size of the data structure, the query time and the stretch guarantee.

Awerbuch \etal \cite{AwBeCoPe98} presented a distance oracle with stretch $64k$, $\tilde{O}(k n^{1+1/k})$ size, $\tilde{O}(m n^{1/k})$ construction time and $O(k n^{1/k})$ query time.
Cohen \cite{Coh98} improved the stretch to $2k+\eps$ and later Matou{\v{s}}ek \cite{Ma96} further improved the stretch to $2k-1$ using completely different techniques.

In a seminal paper Thorup and Zwick \cite{ThZw05} significantly improved the query time, presenting a distance oracle with
$2k-1$ stretch, $O(k n^{1+1/k})$ expected size, $O(k m n^{1/k})$ construction time and $O(k)$ query time.
The stretch-space tradeoff is essentially optimal up to the $k$ factor in the space, assuming the girth conjecture of Erd\H{o}s \cite{Er64}.
Roditty, Thorup, and Zwick \cite{RTZ05} later show how to de-randomize the construction while keeping the same bounds.
Baswana and Kavitha \cite{BaKa06} presented an improved construction time for dense graphs of $O(n^2 \log{n})$ with query time of $O(k)$ for $k>2$ and of $\Theta(\log{n})$ for $k=2$.

Wulff-Nilsen \cite{Wu12} presented a distance oracle with subquadratic time when $m=o(n^2)$, presenting a distance oracle with $2k-1$ stretch, $O(k n^{1+1/k})$ size, $O(k)$ query time and $O(\sqrt{k}m + kn^{1+c/\sqrt{k}})$ construction time for some absolute constant $c$.

P{\v a}tra{\c s}cu and Roditty \cite{PaRo10} presented a distance oracle for unweighted graphs of size $O(n^{5/3})$ with a multiplicative stretch 2 and additive stretch 1. In addition, they present a distance oracle for weighted graphs of size $O(n^{2}/\sqrt[3]{\alpha})$ where $\alpha = n^2/m$.

Later Abraham and Gavoille \cite{AG11} presented a distance oracle of size $\tilde{O}(n^{1+2/(2k-1)})$ with $O(k)$ query time and with a multiplicative stretch $2k-2$ and additive stretch 1.

Baswana \etal \cite{BaGaSeUp08} also considered distance oracles with both additive and multiplicative stretch, presenting a distance oracle of size $O(k n^{1+1/k})$,
with a multiplicative stretch $2k-1$ and additive stretch 2 with subquadratic construction time of $O(\min(m+kn^{3/2+1/(2k)+1/(2k-2)}, kmn^{1/k}))$.

Mendel and Naor \cite{MN06,MN07} studied approximate distance oracles with constant query time.
They presented an approximate distance oracle with size $O(n^{1+1/k})$, $128k$ stretch, $O(1)$ query time and $O(n^{2+1/k} \log{n})$ construction time.
The $128k$ stretch can be improved to $33k$ using techniques of Naor and Tao \cite{NT12}, and according to Naor and Tao the stretch can be further improved to $16k$ using a more careful analysis. In addition, Mendel and Schwob \cite{MeSc09} improved the $O(n^{2+1/k} \log{n})$ construction time to $O(mn^{1/k} \log^3{n})$.

Wulff-Nilsen \cite{Wu13} improved the $128k$ stretch of Manor and Naor's construction to $(2k+\eps)$ at the cost of additional $k$-factor in the size, the query time
of his construction is $O(1/\eps)$ and the construction time is
$O(kmn^{1/k} + kn^{1+1/k} \log{n} + mn{1/(ck)} \log^3{n})$.
For the case of $k = O(\log{n}/\log\log{n})$ and a fixed $\eps$, Wulff-Nilsen showed that it is actually possible to reduce the size back to
$O(n^{1+1/k})$.
In addition, Wulff-Nilsen \cite{Wu13} showed that it is possible to improve the query time of Thorup and Zwick's distance oracle \cite{ThZw05} from $O(k)$ to $O(\log{k})$.
Namely, a distance oracle of size $O(k n^{1+1/k})$, $2k-1$ stretch, $O(\log{k})$ query time.

In this paper we further improve the $O(\log{k})$ query time of Wulff-Nilsen's distance oracle \cite{Wu13} from $O(\log{k})$ to a universal constant.
More precisely, we show a distance of size $O(k n^{1+1/k})$, $2k-1$ stretch, $O(1)$ query time and $O(kmn^{1/k} + kn^{1+1/k} \log{n} + mn{1/(ck)} \log^3{n})$ construction time.

%We show how to improve the query time of the Thorup-Zwick oracle to constant while maintaining the $2k-1$ stretch.
Our algorithm first invokes the Mendel and Naor \cite{MN07} distance oracle to obtain an initial distance estimation.
We then show that using this distance estimation it is possible to get $2k-1$ stretch in constant time using the Thorup-Zwick distance oracle and some additional information.

%The girth conjecture has been proved for the specific cases of $k=1,2,3$, and $5$ \cite{Wen91}.

%We improve the long standing state of the art for compact routing schemes for general graphs.
%We show the existence of routing schemes that use ${\tilde O}(n^{1/k})$ tables size with $c\cdot k$ stretch for $c <4$.
%This is the first improvement to the stretch-space tradeoff of routing scheme since the result of Thorup and Zwick [SPAA'01].
%As part of our techniques, we present new important properties on the Thorup-Zwick construction, which may be of independent interest.

%\paragraph{Paper Organization:}

%
%To simply the presentation, we focus on the case where $k=4$ and present a routing scheme with stretch 10.52.
%In Section \ref{sec:any-k}, we present our construction for general $k$.
%Due to space limitations some of the proofs are deferred to the appendix.

%==========================================================================
%\section{Outline and main ideas}

%\section{Preliminaries and Notation}

%---------------------------------------
\subsection{Preliminaries}

For completeness we first outline the construction of Thorup-Zwick distance oracle \cite{ThZw05}.

For a given positive integer $k$, the sets $V = A_0 \supseteq A_1 \supseteq \cdots \supseteq A_{k-1}$ are constructed as follows.
Set $A_0 \gets V$ and $A_k \gets \emptyset$.
The set $A_i$ for $1< i \leq k-1$ is obtained by sampling the vertices of $A_{i-1}$ independently at random with probability $n^{-1/k}$.

For pair of nodes $u$ and $v$, let $\dist(u,v)$ be their distance in $G$.
The pivot $p_i(v)$ is defined to be the closest node to $v$ in $A_i$ (break ties arbitrarily).

The bunch of $v$ is defined as follows,
$$B(v) = \bigcup_{i=0}^{k-1}{B_i(v)},$$ where
$$B_i(v) = {\{u \in A_i \setminus A_{i+1} \mid \dist(v,u) < \dist(v,p_{i+1}(v))\}}.$$

The query algorithm given pair of nodes $s$ and $t$ is done as follows.
Let $w \gets s, j\gets 0$.
While $w \notin B(t)$ do the following.
Set $j \gets j+1$, $(s,t) \gets (t,s)$ and $w \gets p_j(s)$.
In the end of the while loop return $\dist(s,w) + \dist(w,t)$.

It was shown in \cite{ThZw05} that the expected size of each bunch is $O(k \cdot n^{1/k})$ and that all bunches can be constructed in $O(kmn^{1/k})$ time.

%==========================================================================
\section{Oracle with Constant Query Time}

Let us start with some notations.
Consider the sets $A_0,...,A_k$ from Thorup-Zwick distance oracle \cite{ThZw05}.
Let $\delta_j(u) = \dist(u,p_{j}(u)) - \dist(u,p_{j-2}(u))$.
Let $\Delta_j(u) = \max\limits_{1\leq i \leq j, ~i ~is~even}{\delta_i(u)}$.

Let $I(i,u)$ be the even index such that  $I(i,u) \leq i$ and $\delta_{I(i,u)}(u)$ is maximal, i.e., $\delta_{I(i,u)}(u) = \Delta_i(u)$

\begin{definition}[Legitimate Pair]
We say that two indices $i_1$ and $i_2$ are legitimate pair for the pair $s,t$ if the following holds:
\beginsmall{enumerate}
\item
$i_2$ and $i_1$ are even.
%\item
%$i_1 < i_2-2$.
\item
$\Delta_{i_1}(s)/2 \leq \dist(s,t)$.
\item
$p_{I(i_2-2,s)}(s) \in B(t)$ or $p_{I(i_2-1,s)}(t) \in B(s)$.
\item
$\dist(s,p_{i_2}) \leq 2\dist(s,p_{i_1})$.
\endsmall{enumerate}
\end{definition}

%5. $\Delta_{i_2}(s) \leq c \Delta_{i_1}(s)$, for some constant $c \leq 128$.

%For every $i$ and $u$ store $I(i,u)$.

Our algorithm consists of two parts.
The first part finds legitimate indices $i_1$ and $i_2$, the second part uses these indices to get $2k-1$ stretch.
In Section \ref{sec:leg} we describe the preprocessing phase and the query phase for finding legitimate pair for a given pair of vertices $s$ and $t$.
In Section \ref{sec:query} we describe the preprocessing phase and the query phase for estimating the distance within stretch $2k-1$ given pair of nodes $s$ and $t$ and legitimate pair $i_1$ and $i_2$.

%==========================================================================
\subsection{Finding Legitimate Pair}
\label{sec:leg}

We now turn to describe the preprocessing and query phases of the first part of the algorithm, which finds legitimate pair of indices given pair of nodes $s$ and $t$.
In order to find a legitimate indices $i_1$ and $i_2$ we use the techniques introduced in \cite{Wu13} and adapt them to our needs.

%The construction of the distance oracle is as follows.
Let us start with describing the information stored at the preprocessing phase.

Construct the Thorup and Zwick distance oracle.
Let ${\cal D}_{TZ} = \{\dist(u,v) \mid u,v\in V ~and~ u \in B(v) \} \cup \{\dist(u,p_i(u)) \mid u\in V, 1 \leq i\leq k-1 \}$.
For every node $v$ and index $1\leq j\leq k-1$, store $\Delta_j(u)$ and the index $I(j,u)$.

Construct the Mendel and Naor \cite{MN07} distance oracle.
Let $\dist_{MN}(u,v)$ be the estimated distance returned by the Mendel-Naor distance oracle for the pair $u$ and $v$.
Let ${\cal D}_{MN} = \{\dist_{MN}(u,v) \mid u,v\in V \}$ be the set of all distances that the Mendel-Naor distance oracle can output.

Let ${\cal D} ={\cal D}_{MN} \cup {\cal D}_{TZ}$.

Let $\tilde{{\cal S}}$ be the set obtained by the following process.
Initially set $\tilde{{\cal S}} \gets \emptyset$.
Consider the elements of ${\cal D}$ in a decreasing order, add the current element $x$ to $\tilde{{\cal S}}$ if the previous added element is greater than $2x$.
Let $\tilde{{\cal D}}$ be an array containing all elements of $\tilde{{\cal S}}$ in an increasing order.
In addition store in a hash $\tilde{{\cal H}}$ all elements $\tilde{d}$ in $\tilde{{\cal D}}$, where $\tilde{d}$ is the key and the value is the index $j$ such that $\tilde{{\cal D}}[j] = \tilde{d}$.

It is not hard to verify that the set $\tilde{{\cal D}}$ satisfies the following properties. First, for every element $d \in {\cal D}$ there is an element $\tilde{d}$ in
$\tilde{{\cal D}}$ such that $d \leq \tilde{d} \leq 2d$.
Second, every two consecutive elements $\tilde{d}_1$ and $\tilde{d}_2$ in $\tilde{{\cal D}}$ such that $\tilde{d}_1 < \tilde{d}_2$ satisfy $\tilde{d}_1 \leq \tilde{d}_2/2$.

For every distance $d \in {\cal D}$ store (in a hash) a pointer to the distance $\tilde{{\cal D}}(d) \in \tilde{{\cal D}}$ such that
$\tilde{{\cal D}}(d)$ is the minimal value in $\tilde{{\cal D}}$ such that $d\leq \tilde{{\cal D}}(d)$.
For a value $d'$ such that $d' \notin {\cal D}$, $\tilde{{\cal D}}(d')$ is undefined.

For every node $u$ and index $i$, let $\tilde{d}(i, u)$ be the minimal element in $\tilde{{\cal D}}$ such that $\dist(u,p_i(u)) \leq \tilde{d}(i, u)$.
For every node $u$, store the distances $\tilde{d}(i, u)$ in a hash $H_u$, where the key is $\tilde{d}(i, u)$ and the value is $i$.
In addition store the values $\tilde{d}(i, u)$ in a sorted array $L_u$ in an increasing order.
Store pointers from every index $1\leq i\leq k-1$ to the element $\tilde{d}(i, u)$ in $L_u$.
%For distance $d'$ in $L_u$, store the index $Ind(d',u)$ such that $\tilde{d}(Ind(d',u), u) =d'$.

%To simplify presentation (to avoid handling end cases separately) add to the the end of $L_v$ the following.
%Let $d'$ be the last value in $L_v$.
%If $k-1$ is even then add one additional value $d''$ at the end of $L_v$, set $\overline{even}_u(d'') = k+1$.

%If $k-1$ is odd  then add two additional values $d''$ $d'''$ at the end of $L_v$, set $\overline{even}_u(d'') = k$ and $\overline{even}_u(d''') = k+1$.

Let $\tilde{{\cal D}}[\tilde{d}, i]$ be the $i$'th element in $\tilde{{\cal D}}$ after $\tilde{d}$ or undefined in case $\tilde{d}$ is not in $\tilde{{\cal D}}$.
Namely, let $j$ be the index such that $\tilde{{\cal D}}[j] = \tilde{d}$, then $\tilde{{\cal D}}[\tilde{d}, i] = \tilde{{\cal D}}[j+i]$.
If $j+i < 0$ then return the first element of $\tilde{{\cal D}}$.
Similarly, if $\tilde{{\cal D}}$ contains less than $j+i$ elements return the last one.

Similarly, we define $L_u[\tilde{d},i]$, namely, $L_u[d,i]$ is the $i$'th element in $L_u$ after $\tilde{d}$ or undefined in case $\tilde{d}$ is not in $L_u$.
Namely, let $j$ be the index such that $L_u[j] = \tilde{d}$, then $L_u[\tilde{d}, i] = \tilde{{\cal D}}[j+i]$.
If $j+i < 0$ then return the first element in $L_u$.
Similarly, if $L_u$ contains less than $j+i$ elements return the last one.

Note that $\tilde{{\cal D}}[\tilde{d}, i]$ (and similarly $L_u[\tilde{d}, i]$) can be accesses in $O(1)$ time.
This can be done as follows. Recall that for every distance $\tilde{d}$ that belongs to $\tilde{{\cal D}}$ the algorithm stores (in a hash) the index $j$ such that
$\tilde{{\cal D}}[j] = \tilde{d}$. Hence the index $j$ can be retrieved in constant time. It is not hard to see now that the value $\tilde{{\cal D}}[\tilde{d}, i] = \tilde{{\cal D}}[j+i]$ can be retrieved in constant time.

For every $\tilde{d} \in L_u$ and node $u$, let $\overline{even}_u(\tilde{d})$ be the even index $i$ with maximal $\dist(u,p_i(u))$ such that
$\tilde{{\cal D}}( \dist(u,p_i(u))) = \tilde{d}$ and let $\underline{even}_u(\tilde{d})$ be the minimal.

We now turn describing how to find legitimate indices in the query phase.

Let us first describe a procedure that will be used several times in the query algorithm.
Procedure $\checkind(s,t,i)$ gets an even index $i$ and nodes $s$ and $t$ and returns a distance $\hat{d}$ that satisfies the following.
1. Either $\hat{d} = -1$, $i< k-2$, and $\dist(s,t) \geq \Delta_i(s)/2$, or
2. $\dist(s,t) \leq \hat{d} \leq (2k-1) \dist(s,t)$, or
3. $\dist(s,t) \leq \hat{d} \leq 2 \dist(s,p_{i-2}(s)) + 3\dist(s,t)$.

Procedure $\checkind(s,t,i)$ operates as follows.
Let $j = I(s,i)$.
If $p_{j-2}(s) \in B(t)$ return $\dist(s,p_{j-2}(s)) + \dist(t,p_{j-2}(s))$.
Else if $p_{j-1}(t) \in B(s)$ return $\dist(s,p_{j-1}(t)) + \dist(t,p_{j-1}(t))$.
%
%Else if $p_{j}(s) \in B(t)$ return $\dist(s,p_{j}(s)) + \dist(t,p_{j}(s))$.
%Else if $p_{j+1}(t) \in B(s)$ return $\dist(s,p_{j+1}(t)) + \dist(t,p_{j+1}(t))$.
%
%...
Else if $i =k-2$ then if $p_{i}(s) \in B(t)$ return $\dist(s,p_{i}(s)) + \dist(t,p_{i}(s))$, else return $\dist(s,p_{k-1}(t)) + \dist(t,p_{k-1}(t))$.
Else if $i =k-1$ then   return $\dist(s,p_{k-1}(s)) + \dist(t,p_{k-1}(s))$.
Else return $-1$.

%The query algorithm for given nodes $s$ and $t$ is as follows.
To find a legitimate pair $i_1,i_2$ in the query algorithm for given nodes $s$ and $t$ do the following.
First invoke the Mendel and Naor \cite{MN07} distance oracle to obtain an initial estimation $\dist_{MN}(s,t)$.
Recall that $\dist(s,t) \leq \dist_{MN}(s,t) \leq 128 k \dist(s,t)$.
Let $\tilde{d} = \tilde{{\cal D}}(\dist_{MN}(s,t))$.
The simplest case is when either $s \in B(t)$ or  $t \in B(s)$.
In this case the algorithm can extract the exact distance $\dist(s,t)$ in constant time.
In this case the algorithm returns $(-1,-1,\dist(s,t))$.

Otherwise, an attempt is made to find the maximal value $\tilde{d}_{min}$ in $L_s$ (or $L_t$) such that
$\tilde{d}_{min} \leq  \dist_{MN}(s,t)/256$.

This is done as follows.
Set $i = 2$ and $found1  \gets false$.
While $(i> -9)$ and $(found1 =false)$ do the following.
Let $\tilde{d}_{curr} = \tilde{{\cal D}}[\tilde{d}, i]$.
Check if $H_s(\tilde{d}_{curr} ) \neq null$ or $H_t(\tilde{d}_{curr}) \neq null$ then $found1 = true$.
Else $i = i-1$.
By the end of the while loop, check if $(found1 =false)$, if so return $(-1,-1,2 \tilde{d}_{curr})$.
Otherwise assume w.l.o.g. that $H_s(\tilde{d}_{curr}) \neq null$ (otherwise switch $s$ and $t$).
Move down the list $L_s$ from the value $\tilde{d}_{curr}$ until finding a value
$\tilde{d}_{min}$ such that $\tilde{d}_{min} \leq \dist_{MN}(s,t)/256$.
Let $i_{min} \gets \overline{even}_s(\tilde{d}_{min})$.
Let $\ell = 0$.
Move up the list $L_s$ until finding a value $L_s[\tilde{d}_{min},\ell]$ such that
$\checkind(s,t,i) \neq -1$ where $i = \overline{even}_s(L_s[\tilde{d}_{min},\ell])$.
This is done until either finding such a value $L_s[\tilde{d}_{min},\ell]$ or until $L_s[\tilde{d}_{min},\ell-2] > \tilde{d}$.
If no such value found return $(-1,-1,\dist_{MN}(s,t))$.
Else let $i_2 \gets I(s, \overline{even}_s(L_s[\tilde{d}_{min},\ell]))$
and $i_1 \gets \underline{even}_s(L_s[\tilde{d}_{min},\ell])$.
If $\checkind(s,t,i_1) \neq -1$, then return $(-1,-1,\checkind(s,t,i_1))$.
Else return $(i_1,i_2, -1)$.

See Procedure \findleg~ for the pseudo-code.

%Let $\tilde{{\cal D}}$ be an array and let $\tilde{I}(d)$ be the index of $$

\begin{figure}[ht]
\begin{center}
\framebox{\hspace{1cm}\parbox{6in}{
{\tt algorithm}$\;$ \findleg$(G)$ $\phantom{2^{2^{2^2}}}$ \\[7pt]
\begin{enumerate}
\item
Let $\tilde{d} = \tilde{{\cal D}}(\dist_{MN}(s,t))$.
\item
If $s \in B(t)$ or  $t \in B(s)$ then return $(-1,-1,\dist(s,t))$.
\item
Set $i = 2$ and $found1  \gets false$.
\item
While $(i> -9)$ and $(found1 =false)$ do:
\begin{enumerate}
\item
Let $\tilde{d}_{curr} = \tilde{{\cal D}}[\tilde{d}, i]$.
\item
If $H_s(\tilde{d}_{curr} ) \neq null$ or $H_t(\tilde{d}_{curr}) \neq null$ then $found1 = true$.
\item
Else $i = i-1$.
\end{enumerate}
\item
If $found1 =false$ then return $(-1,-1,2 \tilde{d}_{curr})$.
\item
If $H_s(\tilde{d}_{curr}) = null$ then set $(s,t) \gets (t,s)$.
\item
Set $j = 0$.
\item
While $(L_s[\tilde{d}_{curr},j] > \dist_{MN}(s,t)/256 )$
\begin{enumerate}
\item
$j \gets j-1$.
\end{enumerate}
\item
$\tilde{d}_{min} \gets L_s[\tilde{d}_{curr},j]$.
\item
Let $i_{min} \gets \overline{even}_s(\tilde{d}_{min})$.
\item
If $\checkind(s,t,i_{min}) \neq -1$, then return $(-1,-1,\checkind(s,t,i_{min}))$.
\item
$\ell \gets 0, found2 \gets false$
\item
While $(found2=  false)$ and $(L_s[\tilde{d}_{min},\ell -2] <  \tilde{d})$
\begin{enumerate}
\item
Let $i \gets \overline{even}_s(L_s[\tilde{d}_{min},\ell])$.
\item
$d = \checkind(s,t,i)$.
\item
If $d\neq -1$ then $found2 \gets  true$ else $\ell \gets \ell+1$.
\end{enumerate}
\item
If  $found2 = false$ then return $(-1,-1,\dist_{MN}(s,t))$.
\item
$i_2 \gets I(s, \overline{even}_s(L_s[\tilde{d}_{min},\ell]))$.
\item
$i_1 \gets \underline{even}_s(L_s[\tilde{d}_{min},\ell])$.
\item
If $\checkind(s,t,i_1) \neq -1$, then return $(-1,-1,\checkind(s,t,i_1))$.
\item
Else return $(i_1,i_2, -1)$.
\end{enumerate}
}\hspace{1cm}}
\end{center}
\caption{\label{findleg} Estimate the distance}
\end{figure}

Let us now turn to the analysis of the Procedure \findleg.

We start by an auxiliary claim.
\begin{claim}
\label{lem:gap}
Consider an even index $i$ such that $2\leq i \leq k-1$, one of the following happens:
either 1. $\dist(s,t) \leq \delta_i(s)/2$ or 2. $p_{i-2}(s) \in B(t)$ or 3. $p_{i-1}(t) \in B(s)$.
\end{claim}
\Proof
If either $p_{i-2}(s) \in B(t)$ or $p_{i-1}(t) \in B(s)$ we are done, so assume
$p_{i-2}(s) \notin B(t)$ and $p_{i-1}(s) \notin B(s)$.

Note that by definition of $B(s)$ and $B(t)$ we have $\dist(s, p_{i}(s)) \leq \dist(s, p_{i-1}(t))$ and
$\dist(t, p_{i-1}(t)) \leq \dist(t, p_{i-2}(s))$.

$\dist(s, p_{i}(s)) \leq \dist(s, p_{i-1}(t)) \leq \dist(s,t) + \dist(t, p_{i-1}(t)) \leq \dist(s,t) + \dist(t, p_{i-2}(s)) \leq 2\dist(s,t) + \dist(s, p_{i-2}(s))$.
We get that $\dist(s, p_{i}(s)) - \dist(s, p_{i-2}(s)) \leq 2\dist(s,t)$. Hence $\dist(s,t) \geq \delta_i(s)/2$, as required.
\QED

We now turn to the correctness of Procedure $\checkind$.

\begin{lemma}
\label{lem:checkind}
Procedure $\checkind(s,t,i)$ gets an even index $i$ and nodes $s$ and $t$ and returns a distance $\hat{d}$ that satisfies the following.
1. Either $\hat{d} = -1$, $i< k-2$, and $\dist(s,t) \geq \Delta_i(s)/2$, or
2. $\dist(s,t) \leq \hat{d} \leq (2k-1) \dist(s,t)$, or
3. $\dist(s,t) \leq \hat{d} \leq 2 \dist(s,p_{i-2}(s)) + 3\dist(s,t)$.
\end{lemma}
\Proof
Let $j = I(s,i)$.

If $p_{j-2}(s) \in B(t)$ then the Procedure returns $\dist(s,p_{j-2}(s)) + \dist(t,p_{j-2}(s))$.
Note that $\dist(s,p_{j-2}(s)) + \dist(t,p_{j-2}(s)) \leq \dist(s,p_{j-2}(s)) + \dist(s,p_{j-2}(s)) + \dist(s,t) =
2\dist(s,p_{j-2}(s)) + \dist(s,t) < 2\dist(s,p_{j-2}(s)) + 3\dist(s,t) \leq 2\dist(s,p_{i-2}(s)) + 3\dist(s,t)$.

Consider now the case where $p_{j-2}(s) \notin B(t)$ and $p_{j-1}(t) \in B(s)$.
In this case the algorithm returns $\dist(s,p_{j-1}(t)) + \dist(t,p_{j-1}(t))$.
Note that since $p_{j-2}(s) \notin B(t)$, we have $\dist(t,p_{j-1}(t)) \leq \dist(t,p_{j-2}(s)) \leq \dist(s,p_{j-2}(s)) + \dist(s,t)$.
Hence, $\dist(s,p_{j-1}(t)) + \dist(t,p_{j-1}(t)) \leq 2\dist(t,p_{j-1}(t)) + \dist(s,t) \leq 2\dist(s,p_{j-2}(s)) + 2\dist(s,t) + \dist(s,t) =
2\dist(s,p_{j-2}(s)) + 3\dist(s,t) \leq 2\dist(s,p_{i-2}(s)) + 3\dist(s,t)$.

Assume $p_{j-2}(s) \notin B(t)$ and $p_{j-1}(t) \notin B(s)$.
In this case we have $\dist(s, p_{j}(s)) \leq \dist(s, p_{j-1}(t)) \leq \dist(s,t) + \dist(t, p_{j-1}(t)) \leq \dist(s,t) + \dist(t, p_{j-2}(t)) \leq 2\dist(s,t) + \dist(s, p_{j-2}(t))$.
We get that $\dist(s, p_{j}(s)) - \dist(s, p_{j-2}(t)) \leq 2\dist(s,t)$. Hence $\dist(s,t) \geq \Delta_i(s)/2$.

Consider now the case where $i =k-2$ and $p_{i}(s) \in B(t)$, in this case the algorithm returns $\dist(s,p_{i}(s)) + \dist(t,p_{i}(s))$.
Note that $\dist(s,p_{i}(s)) \leq i \Delta_i(s)/2$.
We get that $\dist(s,p_{i}(s)) + \dist(t,p_{i}(s)) \leq 2\dist(s,p_{i}(s)) +\dist(s,t) \leq 2i \Delta_i(s)/2 +\dist(s,t) \leq 2(k-2)\Delta_i(s)/2 +\dist(s,t) \leq (2k-3)\dist(s,t)$.

Consider the case where
$i =k-2$, $p_{i}(s) \notin B(t)$ and the algorithm returns $\dist(s,p_{k-1}(t)) + \dist(t,p_{k-1}(t))$.
We get that $\dist(s,p_{k-1}(t)) + \dist(t,p_{k-1}(t)) \leq 2\dist(t,p_{k-1}(t)) + \dist(s,t) \leq 2\dist(t,p_{k-2}(s)) + \dist(s,t)\leq (2k-1)\dist(s,t)$.

The case where $i =k-1$ and the algorithm returns $\dist(s,p_{k-1}(s)) + \dist(t,p_{k-1}(s))$ can be proved similarly to the previous case.

Finally, consider the last case where the algorithm returns -1.
Note that in this case $\dist(s,t) \geq \Delta_i(s)/2$ and $i < k-2$ as required.
\QED

We are now ready to prove the correctness of Procedure \findleg.

\begin{lemma}
Let $(i_1,i_2, \hat{d})$ be the tuple returned by Procedure \findleg, then either $\hat{d} =-1$ and $(i_1,i_2)$ is a legitimate pair for $s$ and $t$ or
$\dist(s,t) \leq  \hat{d} \leq (2k-1)\dist(s,t)$.
\end{lemma}
\Proof
Note that the algorithm returns a value in one of the lines: 2,5,11,14,17,18.
We show that the value returned in each such line satisfies the lemma.

Consider the case where the algorithm returns a value in line 2.
Note that if $s \in B(t)$ or $t \in B(s)$ then the exact distance $\dist(s,t)$ can be extracted in $O(1)$ time.

Consider the case where the algorithm returns a value in line 5.
Let $(-1,-1,2 \tilde{d}_{curr})$ be the tuple returned by the algorithm.
The algorithm returns a value in line 5 of the algorithm in case $found1= false$ when the algorithm reaches line 5.
This happens only when none of the values $\tilde{{\cal D}}[\tilde{d}, i]$ for $-9 \leq i \leq 2$ exists in both $H_s$ and $H_t$.
In other words, both $H_s$ and $H_t$ do not contain values between $\tilde{{\cal D}}[\tilde{d}, -9]$ to $\tilde{{\cal D}}[\tilde{d}, 2]$.
This means that both $H_s$ and $H_t$ do not contain values between $\tilde{d}/2^{-9}$ to $4 \tilde{d}$.

We claim that $\tilde{d}_{curr} \leq \tilde{d}/2^{-9}$.
Note that the only case where $\tilde{d}_{curr} > \tilde{d}/2^{-9}$ is when $\tilde{{\cal D}}$ contains less than nine elements that are smaller than $\tilde{d}$.

we claim that both $H_s$ and $H_t$ contains a value smaller or equal to $\tilde{d}$.
To see this note that as $t \in B(s)$, we have $\dist(s, p_1(s)) \leq \dist(s,t) \leq \dist_{MN}(s,t)$.
Hence $\tilde{{\cal D}}(\dist(s, p_1(s))) \leq \tilde{{\cal D}}(\dist_{MN}(s,t)) = \tilde{d}$.

Note also that $\tilde{{\cal D}}(\dist(s, p_1(s)))$ is contained in both $H_s$ and $\tilde{{\cal D}}$.
It follows that by moving down the list of $\tilde{{\cal D}}$ the algorithm either encounters the value $\tilde{{\cal D}}(\dist(s, p_1(s)))$ or it doesn't reach the beginning of the list.
We get that $\tilde{d}_{curr} \leq \tilde{d}/2^{-9}$.
%In addition, we have that $L_s$ contains a value smaller than $\tilde{d}_{curr}$.

Hence $\tilde{d}_{curr} \leq \tilde{d}/2^{-9} \leq 2\dist_{MN}(s,t)/2^{-9} = \dist_{MN}(s,t)/2^{-8} \leq 128 k \dist(s,t)/2^{-8} = k \dist(s,t)/2$.
Hence  $2 \tilde{d}_{curr} \leq k \dist(s,t) < (2k-1)\dist(s,t)$.
We need to show the other direction, namely, $\dist(s,t) \leq 2\tilde{d}_{curr}$.

Let $j$ be the maximal even index such that $\dist(s, p_j(s)) \leq \tilde{d}_{curr}$.
Note that such an index exists as $\dist(s, p_0(s)) = 0 \leq \tilde{d}_{curr}$.
%Recall that as mentioned above $L_s$ contains a value smaller than $\tilde{d}_{curr}$ hence such an index $j$ exists.
Recall also that $H_s$ does does not contain any value in $\tilde{{\cal D}}[\tilde{d}, i]$ for $-9 \leq i \leq 2$.

%In addition,
This implies that either $j+2 > k-1$ or that
$\tilde{{\cal D}}(\dist(s, p_{j+2}(s))) > \tilde{{\cal D}}[\tilde{d}, 2] \geq 4 \tilde{d} \geq 4 \dist(s,t)$.

In the latter case we have $\Delta_{j+2}(s) > 2 \dist(s,t)$.
Hence by claim \ref{lem:gap}, either $p_j(s) \in B(t)$ or $p_{j+1}(t) \in B(s)$.
Note that also in the first case we have either $p_j(s) \in B(t)$ or $p_{j+1}(t) \in B(s)$.

If $p_j(s) \in B(t)$, then note that $\dist(p_{j}(s),t) \leq \dist(s,t) + \dist(s,p_j(s)) < \dist(s,t) + \tilde{d}_{curr} < \tilde{{\cal D}}[\tilde{d}, 2]$.
But recall that, $H_t$ does not contain values between $\tilde{{\cal D}}[\tilde{d}, -9]$ to $\tilde{{\cal D}}[\tilde{d}, 2]$.
It follows that $\dist(t, p_{j}(s)) \leq \tilde{{\cal D}}(\dist(t, p_{j}(s))) \leq \tilde{d}_{curr}$.
Hence $\dist(s, p_{j}(s))+\dist(t, p_{j}(s)) \leq 2\tilde{d}_{curr}$, as required.

The case where $p_{j+1}(t) \in B(s)$ is handled similarly.

Consider the case where the algorithm returns a value in line 11.
In this case, $\checkind(s,t,i_{min}) \neq -1$.
By Lemma \ref{lem:checkind} $\hat{d}$ satisfies one of the following.
Either $\dist(s,t)\leq \hat{d} \leq (2k-1)\dist(s,t)$ or
$\dist(s,t)\leq \hat{d} \leq 2 \dist(s,p_{i_{min}-2}(s)) + 3\dist(s,t)$.
In the first case the lemma holds. Consider the second case.
Note that in this case $\tilde{d}_{min} \leq \dist_{MN}(s,t)/256 \leq \dist(s,t)/2$.
We get that $\dist(s,t) \leq \hat{d} \leq 2 \dist(s,p_{i_{min}-2}(s)) + 3\dist(s,t) \leq 2\tilde{d}_{min} + 3\dist(s,t) \leq
3\dist(s,t) + 2k\dist(s,t)/2 = (k+3)\dist(s,t) < (2k-1)\dist(s,t)$ for any $k\geq 4$, as required.

Consider the case where the algorithm returns a value in line 14.
The algorithm returns a value in line 14 in the case where $found2 = false$ when the algorithm reaches line 14.
Let $\ell'$ be the value of $\ell$ in line 14 of the algorithm.
Note that if $found2 = false$ in line 14, then $L_s[\tilde{d}_{min},\ell' -2] \geq  \tilde{d}$.

Let $i \gets \overline{even}_s(L_s[\tilde{d}_{min},\ell])$.
Notice that $\dist(s,p_i(s)) \geq \dist_{MN}(s,t)$.

We have $\dist(s,t) \geq \Delta_i(s)/2$ and
$\dist_{MN}(s,t) \leq \dist(s,p_i(s)) \leq i \Delta_i(s)/2 \leq i \dist(s,t) < k \dist(s,t) < (2k-1)\dist(s,t)$.

Consider the case where the algorithm returns a value in line 17.

Let $i_1 \gets \underline{even}_s(L_s[\tilde{d}_{min},\ell])$.
Note that $i' \gets \overline{even}_s(L_s[\tilde{d}_{min},\ell-1])$, satisfies,  $i' = i_1-2$.

In addition, note that in this case $\checkind(s,t,i_1) \neq -1$ and  $\checkind(s,t,i_1-2) = -1$.

Recall that by Lemma \ref{lem:checkind} $\checkind(s,t,i_1) \leq 2 \dist(s, p_{i_1-2}) + 3 \dist(s,t) \leq 2(i_1-2)\Delta_{i_1-2}(s)/2+ \dist(s,t) \leq 2(i_1-2) \leq \dist(s,t) + 3\dist(s,t) \leq
(2k-1)\dist(s,t)$.

We are left with the case where the algorithm returns a value in line 18.
In this case we show that the pair $(i_1,i_2)$ returned by the algorithm is a legitimate pair for $s$ and $t$.
%....
%Note that in this case either $H_s[\tilde{d}_{curr}] \neq null$ or $H_t[\tilde{d}_{curr}] \neq null$.
%Assume w.l.o.g. that $H_s[\tilde{d}_{curr}] \neq null$ (otherwise switch $s$ and $t$).
%
%Note that the value $\tilde{d}_{min}$ found in the algorithm is the maximal value in $H_s$ such that $\tilde{d}_{min} \leq \tilde{d}/256$.
%Let $\ell'$ be the value of $\ell$ in line 14 of the algorithm.
%Note that $\ell'$ is the minimal index such that $\ell' \geq 0$ and $\checkind(s,t,i(\ell')) \neq -1$, where $i(\ell) = \overline{even}_s(L_s[\tilde{d}_{min},\ell])$.
%Let $i_2 \gets \overline{even}_s(L_s[\tilde{d}_{min},\ell'])$ and
%$i_1 \gets \underline{even}_s(L_s[\tilde{d}_{min},\ell'])$.

1. $i_2$ and $i_1$ are even by definition of $\overline{even}_s$ and $\underline{even}_s$.

2. As $\checkind(s,t,i_1) = -1$ we have  $\Delta_{i_1}(s)/2 \leq \dist(s,t)$.

3. As $\checkind(s,t,i_2) \neq -1$, we have that either $p_{I(s,i_2)-2}(s) \in B(t)$ or $p_{I(s,i_2)-1}(t) \in B(s)$.

4. Since $\tilde{{\cal D}}(\dist(s,p_{i_1}(s))) = L_s[\tilde{d}_{min},\ell']$ and
$\tilde{{\cal D}}(\dist(s,p_{i_2}(s))) = L_s[\tilde{d}_{min},\ell']$ then $\dist(s,p_{i_2}(s)) \leq 2 \dist(s,p_{i_1}(s))$.

\QED

\begin{lemma}
Procedure $\findleg~$ runs in $O(1)$ time.
\end{lemma}
\Proof
It is easy to verify that all operations in Procedure $\findleg~$ can be done in constant time.
We need to show that the number of iterations in  each of the while loops is constant.

First not that the number of iterations in the while loop in line 3 of Procedure $\findleg~$ is at most 11.

Consider now the while loop in line 7 of Procedure $\findleg~$.
Note that $L_s[\tilde{d}_{curr},0] \leq 2 dist_{MN}(s,t)$ and that $L_s[\tilde{d}_{curr},j'-1] \leq L_s[\tilde{d}_{curr},j']/2$.
It is not hard to see now that the number of iteration in the while loop of line 7 is $O(1)$.

Finally, consider the while loop of line 12.
Recall that $\tilde{d}_{min}$ found in the algorithm is the maximal value in $H_s$ such that $\tilde{d}_{min} \leq \tilde{d}/256$.

Hence $L_s[\tilde{d}_{min},9] \geq \tilde{d}$.
Therefore when $\ell = 10$, we have $L_s[\tilde{d}_{min},\ell-2] \geq \tilde{d}$.
It follows by the condition of the while loop that the number of iterations is at most $O(1)$.

%
% $\ell \gets 0, found2 \gets false$
%\item
%While $(found2=  false)$
%\begin{enumerate}
%\item
%Let $i \gets \overline{even}_s(L_s[\tilde{d}_{min},\ell])$.
%\item
%$d = \checkind(i,s)$.
%\item
%If $d\neq -1$ then $found2=  true$ else $\ell \gets \ell+1$.
%\end{enumerate}

\QED

%==========================================================================
\subsection{Estimate the Distance given Legitimate Pair}
\label{sec:query}

We now present a procedure \estimate~ that given two legitimate indices $i_1$ and $i_2$, returns in constant time a distance within stretch $2k-1$.
% or legitimate indices
%$i_1'$, $i_2$ such that $i_1' < i_2$ and $\Delta_{i_1'}(s) \geq 2 \Delta_{i_1}(s)$.

%The algorithm invokes the procedure with the new indices until the procedure outputs a distance within the desired stretch.
%Since the initial indices satisfy $i_1 \leq 128 i_2$, the number of times the procedure is invokes is at most $\log{128} = 7$.

The algorithm stores in the preprocessing the following information.

For every node $v$ and every even index $1\leq i \leq k-1$, store the following.

The minimal even index $x_1(i,v)$ such that $1\leq x_1(i,v)\leq k-1$  and
$(x_1(i,v)-i) (\Delta_{x_1(i,v)}(v) - \Delta_{i}(v)) \geq (k-x_1(i,v)-2)\Delta_{i}(v)$.

The minimal even index $x_2(i,v)$ such that $x_1(i,v) \leq x_2(i,v) \leq k-1$ and  $(x_2(i,v)-x_1(i,v)) (\Delta_{x_2(i,v)}(v) - \Delta_{x_1(i,v)}(v)) \geq (k-x_2(i,v)-2)\Delta_{x_1(i,v)}(v)$.

The minimal even index $x_3(i,v)$ such that $x_2(i,v)\leq x_3(i,v) \leq k-1$ and
$(x_3(i,v)-x_2(i,v)) (\Delta_{x_3(i,v)}(v) - \Delta_{x_2(i,v)}(v)) \geq (x_1(i,v))(\Delta_{x_2(i,v)}(v)-\Delta_{x_1(i,v)}(v))$, if no such index exists set $x_3(i,v)$ to be the maximum even index (namely, either $k-2$ or $k-1$).  %= k-1$.

%%For every even index $i$ and node $v$, store the index $J(i,v)$ such that $J(i,v)$ is minimal, even and $\delta_{J(i,v)}(v) \geq 2\dist(v,p_{i}(v))/i$.
%%Note that an index $J(i,v) \leq i$ always exists.

Procedure \estimate~ given $i_1$, $i_2$ operates as follows.
%Let $x_1' = x_1(i_1,s)-1$ and $x_1' = I(x_1',s)$.

%If $\checkind(J(i_2,s),s,t) = -1$ then return $$

Let $x_1 = x_1(i_1,s)$, $x_2 = x_2(i_1,s)$ and $x_3 = x_3(i_1,s)$.

%
%If $\checkind(J(i_1,s),s) = -1$ then set $i_1 = J(i_1,s)$.
%Else if $\checkind(J(i_1,s)-2,s) \neq -1$ then set $i_2 = J(i_1,s)-2$.
%Else return $\checkind(J(i_1,s),s)$.

If $\checkind(s,t,x_1) \neq -1$ then return $\checkind(s,t,x_1)$.
Else if $\checkind(s,t,x_2) \neq -1$ then return $\checkind(s,t,x_2)$.
Else if $\checkind(s,t,x_3) \neq -1$ then return $\checkind(s,t,x_3)$.
Else return $\checkind(s,t,i_2)$.

\begin{figure}[ht]
\begin{center}
\framebox{\hspace{1cm}\parbox{6in}{
{\tt algorithm}$\;$ \estimate$(G, i_1, i_2)$ $\phantom{2^{2^{2^2}}}$ \\[7pt]
\begin{enumerate}
%\setcounter{enumi}{-1}
%
%If $\Delta_{i_1} \leq 2 \Delta_{i_2}$ then
%
%$2\dist(s,p_{i_1}) \geq \dist(s,p_{i_2})$.
%
%\item
%If $\checkind(J(i_2,s),s) = -1$ then return $\checkind(i_2,s,t)$.
%\item
%Else if $\checkind(J(i_2,s)-2,s) \neq -1$ then set $i_2 = J(i_2,s)-2$.
%\item
%Else return $\checkind(J(i_2,s),s)$.
\item
Let $x_1 = x_1(i_1,s)$, $x_2 = x_2(i_1,s)$ and $x_3 = x_3(i_1,s)$.
\item
If $\checkind(s,t,x_1) \neq -1$ then return $\checkind(s,t,x_1)$.
\item
If $\checkind(s,t,x_2) \neq -1$ then return $\checkind(s,t,x_2)$.
%\item
%If $x_3 = k$ then return $\dist(s,p_{k-1}(s)) + \dist(t,p_{k-1}(s))$.
\item
If $\checkind(s,t,x_3) \neq -1$ then return $\checkind(s,t,x_3)$.
\item
Else return $\checkind(s,t,i_2)$.
\end{enumerate}
}\hspace{1cm}}
\end{center}
\caption{\label{estimate} Estimate the distance}
\end{figure}

%Else if $\Delta_{x_3} \geq 2 \Delta_{i_1}$, return $i_1' = x_2$, $i_2' = i_2$.
%Else return $\dist(s,p_{i_2}(s)) + \dist(p_{i_2}(s),t)$.

We now turn to the analysis of the algorithm.

We first show that $x_1(i,v)$ and $x_2(i,v)$ are well defined for every index even $1 \leq i \leq k-1$.
Note that $x_3(i,v)$ is well defined as if there is no index that satisfy the inequality then
$x_3(i,v)$ is set to be either $k-2$ or $k-1$.

\begin{claim}
For every node $v$ and even index $1\leq i\leq k-1$, $x_1(i,v)$, $x_2(i,v)$ are well defined.
\end{claim}
\Proof
Let $x_1 = x_1(i,v)$.
Recall that $x_1$ is the the minimal even index such that $(x_1-i) (\Delta_{x_1}(v) - \Delta_{i}(v)) \geq (k-x_1-2)\Delta_{i}(v)$.
We need to show that such an index exists. If $k-1$ is even then note that $x_1' = k-1$ satisfies the inequality.
Similarly, if $k-2$ is even then note that $x_1' = k-2$ satisfies the inequality.
As one of $k-1$ and $k-2$ is even then $x_1$ is well defined.
Similarly, we can show that $x_2(i,v)$ is well defined.
\QED

\begin{lemma}
Given a pair of nodes $s$ and $t$ and a legitimate pair $i_1$ and $i_2$ for $s$ and $t$,
Procedure \estimate ~returns a distance $\hat{d}$ such that $\dist(s,t) \leq  \hat{d} \leq (2k-1)\dist(s,t)$.
\end{lemma}
\Proof
Notice the the algorithm may halt in one of the following lines: (2)-(5).
We consider these four cases (corresponding to the different lines in which the algorithm may halt in) and show that the lemma holds in each such case.
%
%First we claim that both $x_1$ and $x_2$ are well defined.
%
%
%
%we show that $x_1,x_2$ and $x_3$ are well defined.
%Note that both $x_1' = k-2$  and $x_1'' = k-1$ satisfy the inequality and thus $x_1$ is well defined....
%Similarly $x_2$ is well defined.
%
%

The first case is when the algorithm halts in line 2, namely,
when $\checkind(s,t,x_1) \neq -1$.

By Lemma \ref{lem:checkind} either $\checkind(s,t,x_1) \leq (2k-1)\dist(s,t)$ or
$\checkind(s,t,x_1) \leq 2 \dist(s,p_{x_1-2}(s)) + 3\dist(s,t)$. We need therefore to show that $2 \dist(s,p_{x_1-2}(s)) + 3\dist(s,t) \leq (2k-1)\dist(s,t)$.

Note that by the minimality of $x_1$, we have $(x_1-2-i_1) (\Delta_{x_1-2}(s) - \Delta_{i_1}(s)) < (k-x_1+2-2)\Delta_{i_1}(s)$.
Note also that for every even $j\geq i_1$ we have,
$\dist(s,p_{j}(s))  \leq i_1\Delta_{i_1}(s)/2 + (j-i_1) \Delta_{j}(s)/2$.

It follows that
\begin{eqnarray*}
\hat{d} &\leq& 2 \dist(s,p_{x_1-2}(s)) + 3\dist(s,t) \\ &\leq&
2 i_1\Delta_{i_1}(s)/2 + 2 (x_1-2-i_1) \Delta_{x_1-2}(s)/2 + 3 \dist(s,t) \\ &=&
i_1\Delta_{i_1}(s) + (x_1-2-i_1) \Delta_{x_1-2}(s) + 3 \dist(s,t) \\ &\leq&
i_1\Delta_{i_1}(s) + (x_1-2-i_1) (\Delta_{x_1-2}(s) - \Delta_{i_1}(s)) +
(x_1-2-i_1) \Delta_{i_1}(s) + 3 \dist(s,t) \\ &=&
(x_1-2)\Delta_{i_1}(s) + (x_1-2-i_1) (\Delta_{x_1-2}(s) - \Delta_{i_1}(s)) + 3 \dist(s,t) \\ &\leq&
(x_1-2)\Delta_{i_1}(s) + (k-x_1+2-2)\Delta_{i_1}(s) + 3 \dist(s,t) \\ &=&
(k-2)\Delta_{i_1}(s) + 3\dist(s,t) \leq (2k-4)\dist(s,t) + 3\dist(s,t) \\ &=& (2k-1)\dist(s,t).
\end{eqnarray*}
%
%The second case is is when the algorithm halts in line 5, namely,
%when $p_{x_1-2}(s) \notin B(t)$, and $p_{x_1-1}(t) \in B(s)$.
%In this case the algorithm returns $\dist(t,p_{x_1-1}(t)) + \dist(p_{x_1-1}(t),s)$.
%Note that as $p_{x_1-2}(s) \notin B(t)$, we have $\dist(t,p_{x_1-1}(t)) \leq \dist(s,p_{x_1-2}(s)) + \dist(s,t)$.
%
%
%We thus get,
%\begin{eqnarray*}
%\hat{\dist}(s,t) &=& \dist(s,p_{x_1-1}(s)) + \dist(p_{x_1-1}(s),t) \\ &\leq&
%\dist(s,p_{x_1-2}(s)) + \dist(p_{x_1-2}(s),t) + 2\dist(s,t)\\ &\leq&
%(2k-3)\dist(s,t) + 2\dist(s,t) = (2k-1)\dist(s,t).
%\end{eqnarray*}

Note that if the algorithm did not return a value in line 2, namely, if $\checkind(s,t,x_1) = -1$ then by Lemma \ref{lem:checkind},
we have $\dist(s,t) \geq \Delta_{x_1}(s)/2$.

The proof of the case where the algorithm returns a value in step 3 is similar to the previous case (with replacing $x_1$ by $x_2$ and $i_1$ by $x_1$).

We now turn the the case where the algorithm halts in line 4.

In this case we have by Lemma \ref{lem:checkind}, $\dist(s,t) \geq \Delta_{x_2}(s)/2$.

%We claim that in this case we have $\dist(s,t) \geq \Delta_{x_2}(s)/2$.
%
%To see this, recall that as the algorithm doesn't halt in line $(3)$ and $(4)$, thus we have:
%$p_{x_2-2}(s) \notin B(t)$ and $p_{x_2-1}(t) \notin B(s)$.
%Hence $\dist(t,p_{x_2-1}(t)) \leq  \dist(t,p_{x_2-2}(s)) \leq
%\dist(s,p_{x_2-2}(s)) + \dist(s,t)$ and
%$\dist(s,p_{x_2}(t)) \leq  \dist(s,p_{x_2-1}(t)) \leq
%\dist(t,p_{x_2-1}(t)) + \dist(s,t) \leq \dist(s,p_{x_2-2}(s)) + 2\dist(s,t)$.
%It follows that $\dist(s,t) \geq (\dist(s,p_{x_2}(t))- \dist(s,p_{x_2-2}(s)))/2 = \delta_{x_2}/2 = \Delta_{x_2}/2$.

Recall that $x_3$ is the minimal even index such that
$(x_3-x_2) (\Delta_{x_3}(s) - \Delta_{x_2}(s)) \geq (x_1)(\Delta_{x_2}(s)-\Delta_{x_1}(s))$, or the maximum even index if no such index exists.

By the minimality of $x_3$, we have
$(x_3-2-x_2) (\Delta_{x_3-2}(s) - \Delta_{x_2}(s)) < (x_1)(\Delta_{x_2}(s)-\Delta_{x_1}(s))$.
Therefore $(x_3-2-x_2) (\Delta_{x_3-2}(s)) < (x_1)(\Delta_{x_2}(s)-\Delta_{x_1}(s)) + (x_3-2-x_2)\Delta_{x_2}(s)$.
Note that, $\dist(s, p_{x_3-2}(s)) \leq x_1 \cdot \Delta_{x_1}(s)/2 + (x_2-x_1)\Delta_{x_2}(s)/2 + (x_3-2-x_2)\Delta_{x_3-2}(s)/2$.

We get that,
\begin{eqnarray*}
\dist(s, p_{x_3-2}(s)) &\leq& x_1\Delta_{x_1}(s)/2 + (x_2-x_1)\Delta_{x_2}(s                                                                                                                   )/2 + (x_3-2-x_2)\Delta_{x_3-2}(s)/2 \\ &\leq&
x_1\Delta_{x_1}(s)/2 + (x_2-x_1)\Delta_{x_2}(s)/2 + (x_1)(\Delta_{x_2}(s)-\Delta_{x_1}(s))/2 + (x_3-2-x_2\Delta_{x_2}(s))/2 \\ &=&
(x_3-2)\Delta_{x_2}(s)/2
\end{eqnarray*}

Hence $\hat{d} = 2\dist(s,p_{x_3-2}(s)) + 3\dist(s,t) \leq (x_3-2)\Delta_{x_2}(s) + 3\dist(s,t) \leq
2(x_3-2)\dist(s,t) + 3\dist(s,t) = (2x_3-1)\dist(s,t)\leq (2k-1)\dist(s,t)$.
%
%
%We now turn to case $(6)$.
%Note that in this case we have $\dist(p_{x_3-1}(t)) \leq \dist(p_{x_3-2}(s)) + \dist(s,t)$.
%%In addition,
%%$\dist(s, p_{x_3-2}(s)) \leq \dist(s, p_{x_3-2}(s))\leq (x_3-2\Delta_{x_2})/2$.
%
%Hence $\hat{d} = (\dist(t,p_{x_3-1}(t)) + \dist(p_{x_3-1}(t),s), i_1,i_2) \leq 2 \dist(t, p_{x_3-1}(t)) + \dist(s,t)
%\leq 2 (\dist(s, p_{x_3-2}(s)) +\dist(s,t)) + \dist(s,t) =2 \dist(s, p_{x_3-2}(s)) + 3\dist(s,t) \leq 2(x_3-2)\Delta_{x_2}/2+ 3\dist(s,t) \leq
%2(x_3-2)\dist(s,t)+ 3\dist(s,t) = (2x_3-1)\dist(s,t) \leq (2k-1)\dist(s,t)$.

Finally, consider the case where the algorithm returns a value in line 5.
Note that in this case,we have $\dist(s,t) \geq \Delta_{x_3}(s)/2$.
In addition note that $x_3$ is not the maximal even index as otherwise $\checkind(s,t,x_3) \neq -1$.
Hence by definition $x_3$ satisfies $(x_3-x_2) (\Delta_{x_3}(s) - \Delta_{x_2}(s)) \geq (x_1)(\Delta_{x_2}(s)-\Delta_{x_1}(s))$.

We claim that in this case $\Delta_{x_3}(s) \geq 2 \Delta_{i_1}(s)$.

To show this we consider two cases, the first is when $x_1 \leq  (k+i_1-2)/2$ and the second when $x_1 >  (k+i_1-2)/2$.

Consider the first case where $x_1 \leq (k+i_1-2)/2$.
We have,
$(\Delta_{x_1}(s) - \Delta_{i_1}(s)) \geq (k-x_1-2)\Delta_{i}(s)/(x_1-i_1) \geq  \Delta_{i_1}(s) \cdot (k-(k+i_1-2)/2-2)/((k+i_1-2)/2-i_1) = \Delta_{i_1}(s)$.
We get that $\Delta_{x_1}(s) \geq  2\Delta_{i_1}(s)$.
Hence $\Delta_{x_3}(s) \geq \Delta_{x_1}(s) \geq  2\Delta_{i_1}(s)$.

Consider now the second case where $x_1 >  (k+i_1-2)/2$.
%Note that $x_1 >  (k+i_1-2)/2 \geq k/2$

\begin{eqnarray*}
(\Delta_{x_3}(s) - \Delta_{x_2}(s)) &\geq& (x_1)(\Delta_{x_2}(s)-\Delta_{x_1}(s))/(x_3-x_2) \\ &\geq&
(x_1)(k-x_2-2)\Delta_{x_1}(s)/(x_3-x_2)(x_2-x_1) \\ &\geq& (x_1)(x_3-x_2)\Delta_{x_1}(s)/(x_3-x_2)(x_2-x_1) \\ &=&
(x_1)\Delta_{x_1}(s)/(x_2-x_1) \\
&>&
\Delta_{x_1}(s) (k+i_1-2)/2(x_2-(k+i_1-2)/2) \\ &=&
\Delta_{x_1}(s) (k+i_1-2)/(2x_2-(k+i_1-2)) \\&\geq&
\Delta_{x_1}(s) (k+i_1-2)/(2(k-2)-(k+i_1-2)) \\ &=&
\Delta_{x_1}(s) (k+i_1-2)/(k-2-i_1) \\ &\geq&
\Delta_{x_1}(s).
\end{eqnarray*}

We get that, $\Delta_{x_3}(s) \geq \Delta_{x_2}(s) + \Delta_{x_1}(s) \geq 2 \Delta_{i_1}(s)$ as required.

We thus have
\begin{eqnarray*}
\dist(s,t) \leq \hat{d} &\leq& 2 \dist(s,p_{I(i_2,s)-2}(s)) + 3\dist(s,t) \\ &\leq&
2 \dist(s,p_{i_2}(s)) - 2\Delta_{i_2}(s) + 3\dist(s,t) \\ &\leq&
2 \dist(s,p_{i_2}(s)) - \dist(s,t) \\ &\leq&
4 \dist(s,p_{i_1}(s)) - \dist(s,t) \\ &\leq&
2 i_1 \Delta_{i_1}(s) - \dist(s,t) \\ &\leq&
2 i_1 \dist(s,t) - \dist(s,t) \\ &\leq&
2 (k-1) \dist(s,t) - \dist(s,t) \\ &<&
(2 k-1) \dist(s,t).
\end{eqnarray*}
\QED

\subsection{Running time and space}

The analysis of the running time and space is similar to the one presented in \cite{Wu13} and is brought here for completeness.

The Mendel-Naor distance oracle \cite{MN07} can be constructed in $O(n^{2+1/k} \log{n})$ time and requires $O(n^{1+1/k})$ time.

Similar to the construction of Wulff-Nilsen \cite{Wu13}, our construction can use any distance oracle with $O(c \cdot k)$ stretch and $O(1)$ time that can output at most $O(n^{1+1/k})$ different distances, for any constant $c$.
More precisely, our query algorithm can be slightly modified such that given a distance oracle with $O(c \cdot k)$ stretch and $O(1)$ time, the query time is $O(\log{c})$
for some integer $c$. This can be done by modifying Procedure \findleg~and having $O(\log{c})$ iterations in the while loops.
Constructing such a distance oracle with $O(ck)$ stretch and $O(1)$ query time can be done using the construction of
Mendel and Schwob \cite{MeSc09} in $O(mn^{1/ck} \log^3{n})$.

Constructing the bunches and the pivots can be done by the Thorup-Zwick \cite{ThZw05} analysis in $O(k m n^{1/k})$.

As shown in \cite{Wu13}, the set ${\cal D}_{MN}$ contains $O(n^{1+1/k})$ values.
In addition,  the set ${\cal D}_{TZ}$ contains $O(k n^{1+1/k})$ values.
Hence the set ${\cal D}$ contains $O(k n^{1+1/k})$ values.
Sorting the values in ${\cal D}$ and forming the set $\tilde{{\cal D}}$ can be done in $O(k n^{1+1/k} \log{n})$ time.

Finding the values $\tilde{{\cal D}}(d)$ for every distance $d \in {\cal D}$ can be done in $|{\cal D}|$ by traversing in parallel both lists
$\tilde{{\cal D}}$ and ${\cal D}$ in increasing order as follows.
Initially, set $j,i=0$.
While not reaching the end of both lists do the following.
If ${\cal D}[j] \leq \tilde{{\cal D}}[i]$, set $\tilde{{\cal D}}(d) = \tilde{{\cal D}}[i]$, where $d = {\cal D}[j]$ and set $j \gets j+1$.
Else (${\cal D}[j] > \tilde{{\cal D}}[i]$), set $i \gets i+1$.
It is not hard to see that this process takes $O(|{\cal D}|)$ time.

%
%For every distance $d \in {\cal D}$ store (in a hash) a pointer to the distance $\tilde{{\cal D}}(d) \in \tilde{{\cal D}}$ such that
%$\tilde{{\cal D}}(d)$ is the minimal value in $\tilde{{\cal D}}$ such that $d\leq \tilde{{\cal D}}(d)$.
%For values $d' \notin {\cal D}$, $\tilde{{\cal D}}(d')$ is undefined.

Consider a node $u$.
It is not hard to see that finding the values $\delta_j(u)$ for every index $1\leq j\leq k-1$ can be done in $O(k)$ time by simply calculating $\delta_j(u) = \dist(u,p_j(u)) - \dist(u,p_{j-2}(u))$.

Finding the values $\Delta_j(u)$ for every index $1\leq j\leq k-1$ can be done in $O(k \log{k})$ time by the following. First sort the values $\delta_j(u)$ (this takes $O(k \log{k})$ time).
Next, sequentially traverse the indices $i$ from $1$ to $k$ and maintaining the largest value $\delta_j(u)$ observed so far.
Constructing the hash $h_u$ and $L_u$ can also be done in time $O(k)$.

%For every node $u$, finding the values $\Delta_j(u)$ for every index $1\leq j\leq k-1$ can be done in $O(k)$ time by
%sequentially traverse the indices $i$ from $1$ to $k$ and maintaining the largest value $\delta_j(u)$ observed so far.
%Constructing the hash $h_u$ and $L_u$ can also be done in time $O(k)$.

Finally calculating the values $x_1(i,u)$, $x_2(i,u)$, $x_3(i,u)$ can be done in time $O(k^2)$ as follows.

For every pair of indices $i,j$ find
$(j-i) (\Delta_{j}(v) - \Delta_{i}(v))$ and $(k-j-2)\Delta_{i}(v)$

Set $x_1(i,v)$ to be the minimal even index such that $1\leq x_1(i,v)\leq k-1$  and
$(x_1(i,v)-i) (\Delta_{x_1(i,v)}(v) - \Delta_{i}(v)) \geq (k-x_1(i,v)-2)\Delta_{i}(v)$.
This can be done by simply exhaustive search on all indices.

Similarly set  $x_2(i,v)$ to be
the minimal even index such that $x_2(i,v) \leq k-1$ and  $(x_2(i,v)-x_1(i,v)) (\Delta_{x_2(i,v)}(v) - \Delta_{x_1(i,v)}(v)) \geq (k-x_2(i,v)-2)\Delta_{x_1(i,v)}(v)$.

Similarity set  $x_3(i,v)$ to be
the minimal even index  such that $x_3(i,v) \leq k-1$ and
$(x_3(i,v)-x_2(i,v)) (\Delta_{x_3(i,v)}(v) - \Delta_{x_2(i,v)}(v)) \geq (x_1(i,v))(\Delta_{x_2(i,v)}(v)-\Delta_{x_1(i,v)}(v))$, if no such index exists set $x_3(i,v)$ to be the maximum even index (namely, either $k-2$ or $k-1$).  %= k-1$.

We get that the preprocessing time for each node $u$ is $O(k^2) \leq O(k \log{n})$ and thus for all nodes $O(n k \log{n})$ time.

All in all the preprocessing time for given integers $k$ and $c$ is $O(kmn^{1/k} + kn^{1+1/k} \log{n} + mn{1/(ck)} \log^3{n})$.

In addition, it is not hard to verify that the size of the data structure is $O(k n^{1+1/k})$.

%%%%%%%%%%%%%%%%%%%%%%%%%%%%%%%%%%%%%%%%%%%%%%%%

%\bigskip\noindent{\bf Acknowledgement:}
%I'm extremely grateful to Ittai Abraham for very helpful discussions.
%I would like to thank my advisor, David Peleg, for very helpful ideas, comments, and observations. I also thank Michael Langberg and Liam Roditty for useful discussions.

%%%%%%%%%%%%%%%%%%%%%%%%%%%%%%%%%%%%%%%%%%%%%%%%
%\clearpage

\bigskip
%\bigskip
%\bigskip
%\bigskip
%\newpage
%\appendix
%\centerline{\Large\bf Appendix}
%%
%%
%
%
%\section{Some proofs}
%

%\APPENDF
%
%%%%%%%%%%%%%%%%%%%%%%%%%%%%%%%%%%%%%%%%%%%%%%%%

{\small
\begin{thebibliography}{10}

%\bibitem{AbChGaPe10}
%I.~Abraham, S.~Chechik, C.~Gavoille and D.~Peleg.
%\newblock{Forbidden-set distance labels for graphs of bounded doubling dimension.}
%\newblock In {\em In Proc. 29th ACM Symp. on Principles of Distributed Computing (PODC)}, 192--200,  2010.

\bibitem{AG11}
I.~Abraham, and C.~Gavoille.
\newblock On Approximate Distance Labels and Routing Schemes with Affine Stretch.
\newblock {\em DISC}, 404--415, 2011.


\bibitem{AwBeCoPe98}
B.~Awerbuch, B.~Berger, L.~Cowen, and D.~Peleg.
\newblock Near-linear time construction of sparse neighborhood covers.
\newblock In {\em SIAM J. Comput.}, Vol. 28, No. 1, 263-–277, 1998.


%\bibitem{AwKuPe91}
%B.~Awerbuch, S.~Kutten, and D.~Peleg.
%\newblock On buffer-economical store-and-forward deadlock prevention.
%\newblock In {\em Proc. 10th IEEE International Conference on Computer Communications (INFOCOM)}, 410--414, 1991.


\bibitem{BaGaSeUp08}
S.~Baswana, A.~Gaur, S.~Sen and J.~Upadhyay.
\newblock Distance oracles for unwieghted graphs: Breaking the quadratic barrier with constant additive error.
\newblock In {\em ICALP}, pages 609-–621, 2008.


\bibitem{BaKa06}
S.~Baswana and T.~Kavitha.
\newblock Faster algorithms for approximate distance oracles and all-pairs
  small stretch paths.
\newblock In {\em Proc. IEEE Symp. on Foundations of Computer Science (FOCS)}, 591--602, 2006.



%\bibitem{CLPR10}
%S.~Chechik, M.~Langberg, D.~Peleg, and L.~Roditty.
%\newblock $f$-sensitivity distance oracles and routing schemes.
%\newblock In {\em Proc. 8th European Symposium on Algorithms (ESA)},84--96, 2010.

%
%\bibitem{Cohen93}
%E.~Cohen.
%\newblock Fast algorithms for constructing t-spanners and paths with stretch t.
%\newblock In {\em Proc. 34th IEEE Symp. on Foundations of Computer Science (FOCS)},
%  648--658, 1993.


%\bibitem{CT07}
%B.~Courcelle and A.~Twigg.
%\newblock Compact forbidden-set routing.
%\newblock In {\em Proc. 27th international Symposium on Theoretical Aspects of Computer Science (STACS)},  37--48, 2007.

\bibitem{Coh98}
E.~Cohen.
\newblock {Fast algorithms for constructing $t$-spanners and paths with stretch $t$}.
\newblock {\em SIAM J. Comput.}, 28:210–-236, 1998.
%
%\bibitem{Cowen01}
%L.J. Cowen.
%\newblock Compact routing with minimum stretch.
%\newblock {\em J. Alg.}, 38:170--183, 2001.

%\bibitem{BeChKrOv:book}
%M.~de~Berg, O.~Cheong, M.~van~Kreveld and M.~Overmars.
%\newblock Computational Geometry: Algorithms and Applications.
%\newblock {\em Springer}, 2008.


%\bibitem{DuPe10}
%R.~Duan and S.~Pettie.
%\newblock Connectivity oracles for failure prone graphs.
%\newblock In {\em Proc. 42nd ACM Symp. on Theory of Computing (STOC)}, 465--474, 2010.

%\bibitem{EGP03}
%T.~Eilam, C.~Gavoille, and D.~Peleg.
%\newblock Compact routing schemes with low stretch factor.
%\newblock In {\em J. Algorithms}, 46:97--114, 2003.

\bibitem{Er64}
P.~Erd\H{o}s.
\newblock Extremal problems in graph theory.
\newblock In {\em Theory of graphs and its applications}, pages 29–-36, 1964.


%\bibitem{FG95}
%P.~Fraigniaud and C.~Gavoille.
%\newblock Memory requirement for universal routing schemes.
%\newblock In {\em Proc. 14th ACM Symp. on Principles of Distributed Computing (PODC)},   223--230, 1995.
%
%\bibitem{FG01}
%P.~Fraigniaud and C.~Gavoille.
%\newblock Routing in Trees.
%\newblock In {\em 28th Int'l Coll. on Automata, Languages and
% Programming (ICALP)}, 757--772, 2001.
%
%
%\bibitem{GG01}
%C.~Gavoille and M.~Gengler.
%\newblock Space-efficiency for routing schemes of stretch factor three.
%\newblock In {\em J. Parallel Distrib. Comput.}, 61:679--687, 2001.
%
%\bibitem{GP03}
%C.~Gavoille and D.~Peleg.
%\newblock Compact and localized distributed data structures.
%\newblock In {\em Distributed Computing}, 16:111--120, 2003.
%
%\bibitem{GS11}
%C.~Gavoille and  C.~Sommer.
%\newblock Sparse spanners vs. compact routing.
%\newblock In {\em Proc. 23th ACM Symp. on Parallel Algorithms and Architectures
%  (SPAA)}, 225--234, 2011.


\bibitem{Ma96}
J.~Matou{\v{s}}ek.
\newblock On the distortion required for embeding finite metric spaces into normed spaces.
\newblock In {\em Israel Journal of Math} 93, 333--344, 1996.
%  year      = {1996}

%@inproceedings{mat96,
%  author    = {Jiri Matou{\v{s}}ek},
%  title     = {On the distortion required for embeding finite metric
%spaces
%into normed spaces},
%  JOURNAL   = {Israel Journal of Math},
%  VOLUME    = {93},
%  pages     = {333-344},
%  year      = {1996}
%}


\bibitem{MN06}
M.~Mendel,  and A.~Naor.
\newblock Ramsey partitions and proximity data structures
\newblock  In {\em Proc. 47th IEEE Symp. on Foundations of Computer Science (FOCS)}, 109--118, 2006.


\bibitem{MN07}
M.~Mendel,  and A.~Naor.
\newblock Ramsey partitions and proximity data structures.
\newblock In {\em Journal of the European Mathematical Society},
   9:2, 253--275, 2007.

\bibitem{MeSc09}
M.~Mendel and C.~Schwob.
\newblock Fast C-K-R Partitions of Sparse Graphs.
\newblock In {\em Journal of  Theoretical Comp. Sci.}, (2), 1–-18, 2009.


\bibitem{NT12}
A.~Naor, and T.~Tao.
\newblock Scale-oblivious metric fragmentation and the nonlinear Dvoretzky theorem.
\newblock In {\em Israel Journal of Mathematics}, 192 ,489–-504, 2012.

\bibitem{PaRo10}
M.~P{\v a}tra{\c s}cu and L.~Roditty.
\newblock Distance oracles beyond the thorup-zwick bound.
\newblock In {\em FOCS}, pages 815-–823, 2010.


%
%\bibitem{Peleg00}
%D.~Peleg.
%\newblock Distributed computing: a locality-sensitive approach.
%\newblock In {\em SIAM}, 2000.
%
%\bibitem{PeUp89}
%D.~Peleg and E.~Upfal.
%\newblock A trade-off between space and efficiency for routing tables.
%\newblock In {\em J. ACM}, 36(3):510--530, 1989.

%\bibitem{SaKh85}
%N.~Santoro and R.~Khatib.
%\newblock Labelling and implicit routing in networks.
%\newblock In {\em The Computer Journal}, 28(1):5–8, 1985.

\bibitem{RTZ05}
L.~Roditty, M.~Thorup, and U.~Zwick.
\newblock Deterministic constructions of approximate distance oracles and
  spanners.
\newblock In {\em Proc. 32nd Int. Colloq. on Automata, Languages \& Prog.},
  261--272, 2005.

%
%\bibitem{ThZw01}
%M.~Thorup and U.~Zwick.
%\newblock Compact routing schemes.
%\newblock In {\em Proc. 13th ACM Symp. on Parallel Algorithms and Architectures
%  (SPAA)}, 1--10, 2001.

\bibitem{ThZw05}
M.~Thorup and U.~Zwick.
\newblock Approximate distance oracles.
\newblock In {\em J. ACM}, 52, 1--24, 2005.


\bibitem{Wu12}
C.~Wulff-Nilsen.
\newblock Approximate Distance Oracles with Improved Preprocessing Time.
\newblock In {\em Proc. 23rd ACM-SIAM Symposium on Discrete Algorithms (SODA)},202-–208,
2012.


\bibitem{Wu13}
C.~Wulff-Nilsen.
\newblock Approximate Distance Oracles with Improved Query Time.
\newblock In {\em Proc. 24th ACM-SIAM Symp. on Discrete Algorithms (SODA)}, 539--549, 2013.


%\bibitem{Twigg06}
%D.~A.~Twigg.
%\newblock {\em Compact forbidden-set routing}.
%\newblock PhD thesis, University of Cambridge (King's College), 2006.

%\bibitem{VaTa86}
%J.~van~Leeuwen and R.~Tan.
%\newblock Computer networks with compact routing tables.
%\newblock In {\em The Book of} G.~Rozemberg and A.~Salomaa (eds.), 259-–273, 1986.
%
%
%\bibitem{blabla}
%J.~van~Leeuwen and R.~Tan.
%\newblock Computer networks with compact routing tables.
%\newblock In {\em The Book of} G.~Rozemberg and A.~Salomaa (eds.), 259-–273, 1986.

\end{thebibliography}
}
\end{document}